\documentclass[12pt,preprint]{aastex}

\usepackage{amssymb,amsmath,graphicx,color}

\def\be{\begin{equation}}
\def\ee{\end{equation}}

\def\half{{\textstyle{1\over2}}}

\def\half{{\textstyle{1\over2}}}

\def\delcrit{\Delta_{\rm crit}}

\def\au{\textsc{\,au}}

\begin{document}

\title{The statistical mechanics of planet orbits}

\author{Scott Tremaine\altaffilmark{1}}
\affil{Institute for Advanced Study, Princeton, NJ 08540}
\email{tremaine@ias.edu}

\begin{abstract}

\noindent
The final ``giant-impact'' phase of terrestrial planet formation is
believed to begin with a large number of planetary ``embryos'' on
nearly circular, coplanar orbits. Mutual gravitational interactions
gradually excite their eccentricities until their orbits cross and
they collide and merge; through this process the number of surviving
bodies declines until the system contains a small number of planets on
well-separated, stable orbits. In this paper we explore a simple
statistical model for the orbit distribution of planets formed by this
process, based on the sheared-sheet approximation and the ansatz that
the planets explore uniformly all of the stable region of phase
space. The model provides analytic predictions for the distribution of
eccentricities and semimajor axis differences, correlations between
orbital elements of nearby planets, and the complete N-planet
distribution function, in terms of a single parameter, the ``dynamical
temperature'', that is determined by the planetary masses. The
predicted properties are generally consistent with N-body simulations
of the giant-impact phase and with the distribution of semimajor axis
differences in the Kepler catalog of extrasolar planets.  A similar
model may apply to the orbits of giant planets if these orbits are
determined mainly by dynamical evolution after the planets have formed
and the gas disk has disappeared.
\end{abstract}


\section{Introduction}

\label{sec:introd}

\noindent
It is always tempting to apply the powerful tools of statistical
mechanics to macroscopic physical systems that exhibit some degree of
regularity.  The first hint of a role for statistical mechanics in
planet-formation theory came from long-term integrations of the solar
system. These showed that (i) the orbits of all the planets in the
solar system are chaotic, with Liapunov or e-folding times of a few
Myr \citep{sw88,las89,sw92}; (ii) the outer solar system, between
Jupiter and Neptune, is ``full'' in the sense that there are almost no
stable orbits for test particles in this region \citep{hw93,hol97};
(iii) there is a 1--2\% probability that chaotic diffusion of
Mercury's eccentricity will lead to its loss---by ejection, collision
with the Sun, or collision with another planet---within the next 5 Gyr
\citep{lg09}.

In the words of \cite{las96}, these findings lead to the speculation
that ``maybe there was some extra planet at the early stage of
formation of the solar system$\ldots$but this led to so much
instability that one of the planets$\ldots$suffered a close encounter
or a collision with the other ones. This leads eventually to the
escape of the planet and the remaining system gets more stable. In
this case, at each stage, the system should have a time of stability
comparable with its age.'' If this hypothesis is correct, the current
configuration of the solar system and other planetary systems might be
determined, at least in part, by the statistics of orbital
chaos. Moreover this evolution process might be approximately
self-similar, in that the distributions of planet masses,
eccentricities, inclinations, and semimajor axis differences in an
ensemble of planetary systems would remain unchanged except for scale
factors as the systems evolved.

Qualitatively similar ideas have emerged in the exoplanet community,
where they are typically called the ``packed planetary systems
hypothesis''.  The simplest version of this hypothesis
\citep{br04,fm13} is that most planetary systems are ``as tightly
packed as possible'', that is, there is no room for additional planets
on stable orbits. A bold prediction of this hypothesis is that if
there are stable regions of phase space between known exoplanets then
there must be undetected planets in these regions.

Current theories of terrestrial planet formation involve multiple
stages and processes (for recent reviews see \citealt{ki12,mor12,hag13,ray13}). As
the gaseous protoplanetary disk cools, dust condenses and settles into
the disk midplane; the dust particles then accumulate into
planetesimals; and the largest planetesimals undergo runaway growth
until they dominate the gravitational scattering of smaller
planetesimals. At this stage a phase of self-regulated or oligarchic
growth begins, in which the largest planetesimals---now called
planetary embryos---all grow at similar rates. The oligarchic phase
ends when the reservoir of small planetesimals is exhausted. At this
point there are typically a few dozen embryos left. The surviving
embryos gradually excite one another's eccentricities until their
orbits cross and they collide. In this last stage---variously called
late-stage accretion, post-oligarchic growth, or the giant-impact
phase---the number of surviving bodies slowly declines until we are
left with a small number of planets on well-separated, stable
orbits. Laskar's hypothesis or the packed planetary systems hypothesis
are roughly equivalent to the assumption that the giant-impact phase
tends to produce an ensemble of planetary systems with statistically
similar properties, an idea that has been advanced recently from
different perspectives by \cite{mal15}, \cite{pw15}, and \cite{vg15}.

The goal of this paper is to explore a simple model for the
distribution of orbital elements in planetary systems after the
giant-impact phase. The basic ansatz behind the model is that at the
end of the giant-impact phase, the planets explore uniformly all of
the stable phase space that is available to them. This ansatz neglects
many physical processes that may play important roles in the late
stages of planet formation (migration, gas drag, mean-motion and
secular resonances, dynamical friction from a residual planetesimal
population, hit-and-run, fragmenting, and catastrophic collisions,
perturbations from exterior giant planets, etc.). The model does not
describe the distribution of planetary masses following the
giant-impact phase, only the distribution of orbits. We believe the
model is useful because it yields clear predictions, with minimal free
parameters, for several of the observables in multi-planet
systems. It is, of course, no substitute for N-body
simulations, but it can guide our interpretation of these simulations
and does well at reproducing many of their results.

The initial stages of giant planet formation are believed to be
similar to those of terrestrial planets, but the masses of giant
planets are dominated by gas envelopes that are believed to accrete
before the gaseous protoplanetary disk disappears, a few Myr after the
birth of the star. It is possible that the distribution of giant
planets continues to evolve over much longer times as the planets
excite one another's eccentricities. In this case the number of
planets is whittled down mostly by ejection from the system rather
than by collision as in the case of terrestrial planets
\citep{ch08,jt08}\footnote{The different outcomes arise because the
  Safronov number (eq.\ \ref{eq:saf}) is much larger for giant planets
  than terrestrial planets.}. Despite this difference, our ansatz
could also apply to giant planets if this late dynamical evolution is
the primary mechanism that determines the distribution of their
orbits.

\section{The model}

\subsection{The planetary system}

\label{sec:system}

\noindent
We shall work with a simplified model of a planetary system based on
the sheared-sheet or Hill's approximation
\citep{ss53,hen69,hen70,ph86,bt08}. This simplification eliminates
radial gradients in surface density, orbital angular speed, etc., which
otherwise obscure the analysis.

In Hill's approximation we focus on a small radial interval of the
system centered on radius $\bar{a}$ and of width $\Delta a$. Within this interval there are $N$
orbiting masses $m_i$ (``planets'') with semimajor axes $a_i$ and
eccentricities $e_i$. The Hill radius of planet $i$ is
\begin{equation}
r_{Hi}\equiv \bar{a}\left(\frac{m_i}{3M_\star}\right)^{1/3}
\end{equation}
where $M_\star$ is the mass of the host star. The orbital angular
speed is $\Omega_c=(GM_\star/\bar{a}^3)^{1/2}$. We assume that the
planetary orbits are coplanar or nearly so.  We index the planets so
that their semimajor axes are ordered, $ a_1\le a_2\le \cdots\le
a_N$. The positions of the planets are bounded by two fixed radii
$a_0=\bar{a}-\half\Delta a$, $a_{N+1}=\bar{a}+\half\Delta a$, i.e., the
pericenters $a_i(1-e_i)\ge a_0$ and the apocenters $a_i(1+e_i)\le
a_{N+1}$. We ignore any interactions with planets outside this range.

We assume that the planet masses are small, $m_i\ll M_\star$, and
(usually) that the number of planets $N\gg1$. As $N$ grows we assume
that $r_{Hi} \sim \Delta a/N$ since otherwise the dynamics is trivial:
if $r_{Hi}\ll \Delta a/N$ then a system composed of planets on nearly
circular orbits is stable, while if $r_{Hi}\gg \Delta a/N$ there will
be frequent close encounters and collisions so the configuration is
short-lived. This ordering condition can be satisfied either by
letting the planetary masses shrink as $m_i\sim N^{-3}$ for fixed
radial interval $\Delta a$ (Hill's approximation) or by fixing the
planetary masses and letting the radial width grow as $\Delta a\sim N$
(the sheared-sheet approximation).

\subsection{A simplified stability criterion}

\noindent
The criteria for long-term stability of multi-planet systems are not
completely understood. However, N-body integrations of systems of several
planets on nearly circular and coplanar orbits suggest that these systems
are stable if the following approximate stability
criterion is satisfied \citep{cha96,ykm99,zhou07,sl09,funk10,pw15}:
\begin{equation}
a_{i+1}-a_i > \delcrit(t)\, r_{H;i,i+1} \quad\mbox{where}\quad
r_{H;i,i+1}= \frac{a_i+a_{i+1}}{2}\left(\frac{m_i+m_{i+1}}{3M_\star}\right)^{1/3}
\label{eq:stab}
\end{equation} 
is called the mutual Hill radius. This formula assumes that
$a_{i+1}-a_i\ll a_i$ and $m_i+m_{i+1}\ll M_\star$. The stability
parameter $\delcrit(t)$ is a dimensionless function of the age $t$ of
the system in units of the orbital period. Note that most of the
simulations in the literature use equally spaced planets (either in
semimajor axis or log semimajor axis) so they cannot determine whether
(for example) the mean semimajor axis difference or the minimum
semimajor axis difference determines stability. Our stability criterion
(\ref{eq:stab}) {\em assumes} that the minimum separation should be used
in the stability criterion, as would be natural if stability is
determined mostly by interactions with the nearest neighbor. 

The most recent and comprehensive review of numerical determinations
of $\delcrit$ is by \cite{pw15}.  For typical observed exoplanetary
systems, with $t\sim 10^{10}$, they estimate $\delcrit\simeq
10.2$. However, their values are somewhat below those determined by
other studies (see Figure 3 of \citealt{pw15}) so we shall adopt a
slightly more conservative estimate $\delcrit=11\pm 1$. For N-body
simulations, which typically last for $t\sim 10^8$ orbits, we use
$\delcrit=9\pm 1$.

The form of equation (\ref{eq:stab}) suggests that the
minimum stable separation scales with mass as $m^{1/3}$. This
empirical finding is not a consequence of rigorous dynamical
arguments. For example, the resonance overlap criterion
\citep{wis80,deck13} suggests the scaling $m^{2/7}$; however, the
exponents $\frac{1}{3}=0.333$ and $\frac{2}{7}=0.286$ are sufficiently
close that they are hard to distinguish in N-body experiments and the
two scalings yield almost the same results for the purposes of this
paper.  

For eccentric orbits, the most important combination of orbital
elements that determines the stability of adjacent planets is the
distance between the apocenter of the inner orbit and the pericenter
of the outer one \citep{pw15,pet15}. We therefore assume that the system is stable if
\begin{equation}
a_{i+1}-\bar{a}e_{i+1}-a_i-\bar{a}e_i> h_i
\label{eq:stab_ecc}
\end{equation}
where $h_i$ is some function of the masses $m_i$ and $m_{i+1}$. 
This reduces to (\ref{eq:stab}) in the limit of circular orbits if
\begin{equation}
h_i=\delcrit(t)\,\bar{a}\left(\frac{m_i+m_{i+1}}{3M_\star}\right)^{1/3}.
\label{eq:hill}
\end{equation}
An alternative criterion is given by \cite{pet15}, who has conducted
extensive numerical experiments on the stability of systems of two
planets with masses $m_i/M_\star=10^{-4}$--$10^{-2}$. The planets are
on eccentric orbits and are followed for up to $10^8$ orbital
periods. His empirical criterion for stability of closely spaced
planets is equation (\ref{eq:stab_ecc}) with
\begin{equation}
h_i=2.4\bar{a}\left(\frac{\mbox{max\,}(m_i,m_{i+1})}{M_\star}\right)^{1/3}
+ 0.15\bar{a}.
\label{eq:cristobal}
\end{equation}
We normally use equation (\ref{eq:hill}) in our experiments, but we
have also experimented with (\ref{eq:cristobal}) and report results
with both stability criteria. 

The stability criterion (\ref{eq:stab_ecc}) implies that between any pair of adjacent
planets in a stable system there is an excluded length $h_i$.  In a
system of $N$ planets there is a total excluded length
$\sum_{i=0}^{N}h_i$ (for consistency with the assumptions of
\S\ref{sec:system} we take $m_{N+1}=m_0=0$ when determining $h_0$
and $h_N$). Obviously if this sum exceeds $\Delta a$ no stable
planetary system exists. Thus an important parameter is
the filling factor or packing fraction 
\begin{equation}
F=\frac{\sum_{i=0}^N h_i}{\Delta a}, \qquad 0<F<1.
\label{eq:pack}
\end{equation}

\subsection{The distribution of orbits in phase space}

\label{sec:phase}

\noindent
The phase space for coplanar Keplerian orbits has two degrees of
freedom and actions $I_1=(GM_\star a)^{1/2}$, $I_2=(GM_\star
a)^{1/2}[1-(1-e^2)^{1/2}]$. In Hill's approximation the actions can be
taken to be $I_1=\half\Omega_c\bar{a}(a-\bar{a})$ and $I_2=
\half\Omega_c\bar{a}^2e^2$.  The canonical phase-space volume element is
$4\pi^2dI_1dI_2=\pi^2\Omega_c^2\bar{a}^3 da\,de^2$.

We shall assume that the planets are uniformly distributed over the
available phase space, that is, over the phase-space volume that
satisfies the stability criterion (\ref{eq:stab_ecc}). This assumption
is reminiscent of the ergodic hypothesis, which is the basis of much
of equilibrium statistical mechanics, but has a different
interpretation in this case: it would apply, for example, if some
process instantaneously and randomly re-distributed the planets
throughout phase space in an ensemble of planetary systems and then
only the systems having stable configurations survived. This
assumption may well be incorrect: it is more likely that there is a
slow diffusion of planets into the unstable region of phase space,
which would reduce the phase-space density near the stability
boundary. Nevertheless, the ergodic hypothesis is so simple and
powerful that it is worthwhile to explore its implications before
investigating more complex models with more free parameters. We shall
call the models explored here ``ergodic models''. 

Given ergodicity, the $N$-planet distribution function of
semimajor axes ${\bf a}=(a_1,\ldots,a_N)$ and eccentricities ${\bf
  e}=(e_1,\ldots,e_N)$ is
\begin{equation}
dp({\bf a},{\bf e}) =C\,(\pi^2\Omega_c^2\bar{a}^3)^NH(a_1-\bar{a}e_1-a_0-h_0)\prod_{i=1}^N da_ide_i^2\,
H(a_{i+1}-\bar{a}e_{i+1}-a_i-\bar{a}e_i -h_i),
\label{eq:pn}
\end{equation} 
where $H(\cdot)$ is the step function, $C$ is a normalizing
constant, and we set $e_{N+1}=0$.

\subsection{Partition function}

\noindent
The partition function $Z$ is the available volume in
phase space and is given by the integral of the right side of
equation (\ref{eq:pn}) over the semimajor axes and eccentricities,
with $C=1$. To evaluate this integral we first carry out the
integration over semimajor axes. For given eccentricities the total
excluded length is $G=\sum_{i=0}^N h_i+2\bar{a}\sum_{i=1}^N e_i$; the
integral over $\{a_i\}$ depends only on this total and can be written
as
\begin{equation}
  \int_0^{\Delta a-G} \!\!dx_1\int_{x_1}^{\Delta a-G}
    \!\!dx_2\cdots\int_{x_{N-1}}^{\Delta a-G}\!\!dx_N=\frac{1}{N!}H_N(\Delta
      a-G)
\end{equation}
where 
\begin{equation}
H_N(x)=\left\{\begin{array}{ll} x^N &\mbox{if $x>0$} \\ 0 & \mbox{if
    $x\le0$}.
\end{array}
\right.
\end{equation}
Note that $H_0(x)$ is equal to the step function $H(x)$.

After doing the integral over the semimajor axes, the partition
function becomes 
\begin{equation}
Z=\frac{(\pi^2\Omega_c^2\bar{a}^3)^N}{N!}
\prod_{i=1}^N \int de_i^2\,H_N\big(\Delta a-{\textstyle\sum_{i=0}^N}
h_i-2\bar{a}{\textstyle \sum_{i=1}^N }e_i\big).
\label{eq:zzz}
\end{equation} 
In the sheared-sheet approximation, the integrals over $e_i^2$ van be
taken from zero to infinity; then they can be done by induction, giving finally 
\begin{equation}
Z=\frac{\pi^{2N}\Omega_c^{2N}\bar{a}^N(\Delta
  a)^{3N}}{2^N(3N)!\,}(1-F)^{3N}
\label{eq:wo}
\end{equation} 
where the filling factor $F$ is defined in equation
(\ref{eq:pack}). 

This result can be compared to the partition function for a
one-dimensional gas of $N$ hard rods of length $a$ enclosed in a box
of length $L$ \citep{ton36,lieb66}. In this case the filling factor 
is $F=Na/L$ and the partition function is
\begin{equation}
Z=\frac{L^N(1-F)^{N}}{N!}.
\end{equation}
In both cases the partition function depends only on the filling
factor. The main difference between the gas of hard rods and our
model is that the exponent $N$ is replaced by $3N$, reflecting the
extra degrees of freedom arising from the eccentricities.

\subsection{The N-planet distribution function}

\noindent
The $N$-planet distribution function (\ref{eq:pn}) can be rewritten  using the identity 
\begin{equation} 
H(x)=\lim_{\epsilon\to 0^+}\frac{1}{2\pi
  i}\int_{-\infty}^{\infty}\frac{ds}{s-i\epsilon}\exp(isx)
\end{equation} 
(for brevity, in future equations the limit is not written explicitly
but is assumed to apply whenever $\epsilon$ appears). We have 
\begin{align} 
dp({\bf a},{\bf e}) & =\frac{(3N)!\,\bar{a}^{2N}}{2(\pi i)^{N+1}
  (\Delta a)^{3N}(1-F)^{3N}}\int_{-\infty}^\infty
\frac{ds_0}{s_0-i\epsilon}\exp\big[is_0(a_1-\bar{a}e_1-a_0-h_0)\big]\nonumber \\
&\qquad \times \prod_{i=1}^N
 da_i\,de_i^2 \int_{-\infty}^{\infty}
\frac{ds_i}{s_i-i\epsilon}\exp\big[is_i(a_{i+1}-\bar{a}e_{i+1}-a_i-\bar{a}e_i-h_i)\big],
\end{align}
in which we have replaced the normalization constant $C$ by $1/Z$ so
$\int dp({\bf a},{\bf e})=1$, and as usual $e_{N+1}=0$,
$a_0=\bar{a}-\half\Delta a$, $a_{N+1}=\bar{a}+\half\Delta a$. 

The product of exponentials can be re-written as $\exp(i\Phi)$ where
\begin{align}
\Phi & =s_0(a_1-\bar{a}e_1-a_0-h_0)+{\textstyle \sum_{i=1}^N} s_i(a_{i+1}-\bar{a}e_{i+1}-a_i-\bar{a}e_i-h_i) \\
&={\sum_{i=1}^N}\big[a_i(s_{i-1}-s_i)-\bar{a}e_i(s_{i-1}+s_i)-s_ih_i\big]+s_N(\bar{a}+\half
\Delta a)-s_0(\bar{a}-\half \Delta a)-s_0h_0.\nonumber 
\end{align}
The characteristic function of the $N$-planet distribution function is
\begin{equation}
P_N({\bf k},{\bf p})=\int dp({\bf a},{\bf e})\exp\left[i\,{\textstyle \sum_{i=1}^N} (k_ia_i + p_ie_i)\right].
\end{equation}
Carrying out the integrals over $a_i$,
\begin{align}
P_N({\bf k},{\bf p})&=\frac{2^{N-1}(3N)!\,\bar{a}^{2N}}{\pi i^{N+1}(\Delta a)^{3N}(1-F)^{3N}}\int_{-\infty}^\infty
\frac{ds_0}{s_0-i\epsilon}\exp\big[-is_0(\bar{a}-\half\Delta a
+h_0)\big]\nonumber \\
&\quad\times \prod_{i=1}^N \int_{-\infty}^\infty \delta(k_i+s_{i-1}-s_i)
\frac{ds_i}{s_i-i\epsilon}\exp\big[-is_ih_i+i\delta_{iN}s_N(\bar{a}+\half\Delta a)\big] \nonumber \\
&\quad \times \int_0^\infty de_i^2 \exp\big[ie_i(p_i-\bar{a}s_{i-1}-\bar{a}s_i)\big].
\end{align}
Next integrate over $e_i^2$ after multiplying by $\exp(-\epsilon e_i)$ to ensure convergence:
\begin{align}
P_N({\bf k},{\bf p})&=\frac{2^{2N}i^N(3N)! \,\bar{a}^{2N}}{2\pi i(\Delta a)^{3N}(1-F)^{3N}}\int_{-\infty}^\infty
\frac{ds_0}{s_0-i\epsilon}\exp\big[-is_0(\bar{a}-\half\Delta
a+h_0)\big]\nonumber \\
&\quad\times 
\prod_{i=1}^N \int_{-\infty}^\infty \delta(k_i+s_{i-1}-s_i)\frac{ds_i  \exp\big[-is_ih_i+i\delta_{iN}s_N(\bar{a}+\half
  \Delta a)\big]} {(s_i-i\epsilon)(p_i-\bar{a}s_{i-1}-\bar{a}s_i+i\epsilon)^2}.
\label{eq:a1}
\end{align}
This result provides a complete description of the joint probability
distribution of the semimajor axes and eccentricities of all $N$
planets. 

\subsection{The one- and two-planet distribution functions}

\label{sec:onetwo}

\noindent 
For practical purposes, the one- or two-planet distributions are
more useful than the full $N$-planet distribution. Suppose we want
to study the $K$-planet distribution function, starting with
planet $J+1$. Then equation (\ref{eq:a1}) implies that
$k_1=k_2=\cdots=k_J=0$ and this implies that
$s_0=s_1=\cdots=s_J$. Moreover $p_1=p_2=\cdots=p_J=0$. Similarly
$k_{J+K+1}=\cdots=k_N=p_{J+K+1}=\cdots= p_N=0$ and
$s_{J+K}=s_{J+K+1}=\cdots=s_N$. Thus the $K$-planet characteristic
function is 
\begin{align}
&P_K(k_{J+1},\cdots,k_{J+K},p_{J+1},\cdots p_{J+K})\nonumber \\
&\qquad =\frac{2^{2K}i^N(3N)! \,\bar{a}^{2K}}{2\pi i(\Delta a)^{3N}(1-F)^{3N}} \int_{-\infty}^\infty
\frac{ds_J\,\exp\big[is_J(\half\Delta a - \bar{a}-\sum_{j=0}^J h_j)\big]}{(s_J-i\epsilon)^{3J+1}}\\
&\qquad\times \left[\prod_{i=J+1}^{J+K} \int_{-\infty}^\infty
\frac{ds_i  \exp(-is_ih_i)
  \delta(k_i+s_{i-1}-s_i)}{(s_i-i\epsilon)(p_i-\bar{a}s_{i-1}-\bar{a}s_i+i\epsilon)^2}\right]
\frac{\exp\big[is_{J+K}(\bar{a}+\half\Delta a-\sum_{j=J+K+1}^N
  h_j)\big]}{(s_{J+K}-i\epsilon)^{3(N-J-K)}}.\nonumber 
\end{align}

For example let $K=1$. Then 
\begin{align}
P_1(k,p)&=\frac{2i^{N-1}(3N)!\,\bar{a}^2}{\pi(\Delta
  a)^{3N}(1-F)^{3N}}\exp[ik(\bar{a}-\half\Delta a+{\textstyle \sum_{j=0}^J} h_j)] \nonumber \\
&\quad \times \int_{-\infty}^\infty
\frac{ds\exp[is\Delta a(1-F)]}{(s-i\epsilon)^{3(N-J)-2}(s-k-i\epsilon)^{3J+1}(p+k\bar{a}-2s\bar{a}+i\epsilon)^2}.
\end{align}
with $s=s_{J+1}$, $k=k_{J+1}$, $p=p_{J+1}$. The 1-planet distribution
function is the inverse Fourier transform of the characteristic
function,
\begin{align}
p_1(a,e)&=\frac{1}{(2\pi)^2}\int_{-\infty}^\infty dk\,dp\,
\exp[-i(ka+pe)]P_1(k,p)\nonumber \\
&=\frac{4(3N)!\,\bar{a}^2}{(3J)!\,[3(N-J-1)]!\,(\Delta
  a)^{3N}(1-F)^{3N}}H_1(e) H_{3J}\big[a-(\bar{a}-\half\Delta
a+{\textstyle \sum_{j=0}^J}h_j+\bar{a}e)\big]   \nonumber \\
&\quad\times H_{3(N-J-1)}\big[(\bar{a}+\half\Delta a-{\textstyle \sum_{j=J+1}^N}h_j-e\bar{a})-a\big].
\end{align}

If we integrate over semimajor axis, we have 
\begin{equation}
p_1(e)=\int da\, p_1(a,e)= \frac{12N(3N-1)\bar{a}^2}{(\Delta
  a)^2}\frac{(1-F-2e\bar{a}/\Delta a)^{3N-2}}{(1-F)^{3N}}\,e, \qquad F<1.
\label{eq:edist}
\end{equation}
Notice that the result is independent of the planet number
$J$. Moreover it is independent of the excluded lengths $h_i$
associated with the planet in question
or its neighbors; the distribution of eccentricities depends on these lengths
only though the sum of the excluded lengths, which determines the
filling factor $F$ through (\ref{eq:pack}).

Similarly, we can evaluate the two-planet distribution function for
adjacent planets:
\begin{align}
  &p_2(a_{J+1},e_{J+1},a_{J+2},e_{J+2}) \nonumber \\ 
&\quad = \frac{2^4(3N)!\, \bar{a}^4}{(3J)!\,(3N-3J-6)!\,(\Delta a)^{3N}(1-F)^{3N}}
H_1(e_{J+1})H_1(e_{J+2})\nonumber \\
&\qquad \times H_{3J}\big[a_{J+1}-\bar{a}e_{J+1}-(\bar{a}-\half\Delta
a +{\textstyle \sum_{j=0}^J}h_j)\big] \nonumber \\
&\qquad \times H_{3N-3J-6}\big[(\bar{a}+\half\Delta a-{\textstyle \sum_{j=J+2}^N}h_j)-a_{J+2}-\bar{a}e_{J+2}\big]
\nonumber \\
&\qquad \times H_0(a_{J+2}-\bar{a}e_{J+2}-a_{J+1}-\bar{a}e_{J+1}-h_{J+1}\big).
\end{align}

We are interested in the dependence of the two-planet function on
separation $a_{J+2}-a_{J+1}$ so we may integrate over $a_{J+1}$ keeping
the separation fixed. We simplify the notation by setting $a_{J+1}\to a$, $a_{J+2}\to a'$,
$e_{J+1}\to e$, $e_{J+2}\to e'$:
\begin{align}
  &p(a'-a,e,e') \nonumber \\ 
&\quad = \frac{16(3N)!\,\bar{a}^4}{(3N-5)!\, (\Delta a)^{3N}(1-F)^{3N}}
H_1(e)H_1(e')H_0\big[a'-a-\bar{a}(e+e')-h_{J+1}\big]\nonumber\\
&\quad \quad \times H_{3N-5}\big[\Delta a(1-F)
-(a'-a)-\bar{a}(e+e')+h_{J+1}\big].
\label{eq:2df}
\end{align}
Higher-order distribution functions are more complicated but could be
useful; for example, the three-planet distribution function
$p(a,a',a'')$ can be used to describe the distribution of semimajor
axis differences $a''-a$ if an intermediate planet is present but undetected. 

\section{Comparison to simulations and observations}

\label{sec:xxx}

\noindent
We now re-cast the formulas of the preceding sections into simpler forms that
can be more directly compared to numerical simulations, or to
observations. We assume that the number of planets $N$ is large, so
any single system should be considered as a subsystem of one with a
much larger radial width $\Delta a$. This approach is similar to the
grand canonical ensemble of classical statistical mechanics. Formally
we let $N\to\infty$, $\Delta a\sim N$, while the filling factor $F\sim
\mbox{const}$. Thus, for example, a factor like $(1 -x/\Delta a)^N\to
\exp(-Nx/\Delta a)$.

The eccentricity distribution (eq.\ \ref{eq:edist}) is then
\begin{equation}
p_1(e) = \frac{e}{\tau^2}\exp\left(-\frac{e}{\tau}\right),   \qquad \tau\equiv
\frac{\Delta a(1-F)}{6N\bar{a}}.
\label{eq:beta}
\end{equation}
The mean eccentricity is $\langle e\rangle=2\tau$ and we call
$\tau$ the dynamical temperature since it para\-met\-rizes the level of
non-circular motion in the planetary systems. The second of equations
(\ref{eq:beta}) shows that the dynamical temperature is related to the
filling factor and the mean separation $\Delta a/N$; thus it is
determined by the number and masses of the planets. Note also that
the distribution (\ref{eq:beta}) has a fatter tail at high
eccentricities than the standard Rayleigh distribution, $p(e)\propto
e\exp(-\gamma e^2)$.

The two-planet distribution function (\ref{eq:2df}) is
\begin{align}
  p_2(a'-a,e,e') &= \frac{1}{2\bar{a}\tau^5}\,
H_1(e)H_1(e')H_0\big[a'-a-\bar{a}(e+e')-h\big]\nonumber \\
&\quad\times
\exp\left[-\frac{a'-a+\bar{a}(e+e')-h}{2\bar{a}\tau}\right]
\label{eq:twoplanet}
\end{align}
where $h$ is the excluded length between the planets at $a$ and $a'$
(eq.\ \ref{eq:hill}).  Integrating over semimajor axes, the joint
distribution in eccentricities is
\begin{equation}
 p_2(e,e') = \frac{ee'}{\tau^4}\exp\left(-\frac{e+e'}{\tau}\right).
\end{equation}
Thus the distribution of eccentricity of nearest neighbors is
separable: the two-planet eccentricity distribution is the product of
two one-planet distributions (eq.\ \ref{eq:beta}) and there is no
correlation or anti-correlation between the eccentricities of adjacent
planets.

The distribution in semimajor axis difference and total eccentricity
$e_t\equiv e+e'$ is
\begin{align}
p_2(a'-a,e_t)&=\int_0^\infty e\,de\int_0^\infty
e'de'\,\delta(e_t-e-e')p_2(a'-a,e,e') \nonumber \\
&=\frac{e_t^3}{6\bar{a}\tau^5}\,H_0\big(a'-a-\bar{a}e_t-h\big)\exp\left(-\frac{a'-a+\bar{a}e_t-h}{2\bar{a}\tau}\right).
\label{eq:pdaet}
\end{align}

Integrating over the eccentricities gives 
\begin{equation}
  p_2(a'-a)=\frac{4}{3\bar{a}\tau}D\left(\frac{a'-a-h}{2\bar{a}\tau}\right)
\label{eq:da}
\end{equation}
where $D(x)=H(x)[6e^{-x}-e^{-2x}(x^3+3x^2+6x+6)]$. Taking means yields
\begin{equation}
\langle a'-a\rangle=\langle h\rangle +6\bar{a}\tau,
\label{eq:taufind}
\end{equation}
From equation (\ref{eq:pack}), for $N\gg1$ the filling factor can be estimated as 
\begin{equation}
F=\frac{\langle h\rangle}{\langle a'-a\rangle},
\label{eq:ff}
\end{equation}
and with this estimate equation (\ref{eq:taufind}) is equivalent to
the second of equations (\ref{eq:beta}). 

Other authors have examined the relation between the semimajor axis
differences of nearest neighbors and their mutual Hill radius, but
typically by plotting the distribution of $(a'-a)/h$ rather than
$a'-a-h$ \citep{fm13,hm13,ldt14}. In the ergodic model the second of
these has a simpler interpretation: the distribution of $a'-a-h$ is given by equation
(\ref{eq:da}) so long as the dynamical temperature $\tau$ is similar
for all systems in the sample, whereas deriving the distribution of
$(a'-a)/h$ from the ergodic model requires knowledge of the distribution of $h$ as well. 

These formulas give the distribution of eccentricities and semimajor
axis differences for a given value of the free parameter $\tau$. Of
course, different planetary systems may have different values of
$\tau$, so fitting the formulas to a large catalog of planets is only
legitimate if the value of $\tau$ does not vary too much among the
systems in the catalog. To avoid this difficulty, we may plot the
normalized eccentricity,
\begin{equation}
E\equiv \frac{\bar{a}(e'+e)}{a'-a-h},
\label{eq:enorm}
\end{equation}
which has the distribution
\begin{equation}
p(E)=\frac{64E^3}{(1+E)^5}, \qquad 0\le E<1,
\label{eq:enormpred}
\end{equation}
independent of $\tau$.  

If the late stages of terrestrial planet formation are driven by
long-term instabilities and collisions in all planetary systems, and
all collisions result in perfect mergers (as assumed in most
simulations), then we would expect the buildup of planets to proceed
self-similarly, so the final filling factor $F$ would be similar
in all planetary systems. If so, then at a given semimajor axis, age,
and stellar mass $\langle h\rangle\sim m^{1/3}$ (eq.\ \ref{eq:hill}), $\langle
a'-a\rangle\sim m^{1/3}$ (eq.\ \ref{eq:ff}), $\tau\sim
m^{1/3}$ (eq.\ \ref{eq:taufind}), and the surface density
$\Sigma\sim m^{2/3}$ where $m$ is the typical planet mass. 

These scalings fail when the impact velocities become sufficiently
large that the collisions are erosive \citep{sch14}. To explore the
effect of this failure, we parametrize the collisional environment by
the Safronov number,
\begin{equation}
\Theta = \frac{Gm}{R\langle v^2\rangle},
\label{eq:saf}
\end{equation}
where $m$ and $R$ are the planetary mass and radius. Here $\langle
v^2\rangle$ is the mean-square velocity relative to the local circular
speed, equal to $(GM_\star/a)(\frac{5}{8}\langle e^2\rangle +
\frac{1}{2}\langle i^2\rangle)$ where $\langle e^2\rangle$ and
$\langle i^2\rangle$ are the mean-square eccentricity and
inclination. Equation (\ref{eq:beta}) gives $\langle
e^2\rangle=6\tau^2$ so in the simple case where $\langle
e^2\rangle=\langle i^2\rangle$,
\begin{equation}
\Theta = \frac{4m a}{27M_\star R\,\tau^2}.
\end{equation}
For the sample of Kepler planets analyzed in \S\ref{sec:data},
$\langle ma/(M_\star R)\rangle \simeq 0.025$ so
\begin{equation}
\langle\Theta\rangle =
\frac{0.0037}{\tau^2}=\left(\frac{0.06}{\tau}\right)^2.
\label{eq:tauup}
\end{equation}
The typical collision becomes erosive when $v_{\rm esc}/v_\infty
\lesssim 2$--3, where $v_{\rm esc}=(2Gm/R)^{1/2}$ is the escape speed
from the larger body and $v_\infty=2^{1/2}\langle v^2\rangle^{1/2}$ is
the rms relative velocity at infinity (see \citealt{ls12} for a much
more thorough analysis). Thus we expect that the giant-impact phase of
planet formation should have $\langle\Theta\rangle^{1/2}\gtrsim 2$--3,
which in turn implies an upper limit $\tau\lesssim 0.02$--0.03 for the
sample of Kepler planets. These arguments are roughly consistent with
the results of \cite{cha13}, who conducted N-body simulations without
and with fragmentation in collisions and found mean eccentricities
$\langle e\rangle=0.075$ and 0.045 respectively, corresponding to
$\tau=\half\langle e\rangle=0.04$ and 0.02.

As discussed in the Introduction, the approach in this paper is to
model the distribution of orbital elements of a set of planets of
given masses, but not to model how the distribution of planetary
masses is established. This approach is somewhat artificial since
collisions, merging or erosive, affect both the mass and orbit
distributions simultaneously. Nevertheless, the conclusion that the
distribution of planetary masses affects the orbital distribution only
through the dynamical temperature $\tau$ or filling factor $F$ (the
two being related by eq.\ \ref{eq:beta}) appears to be a plausible and
inevitable consequence of the ergodic model.

\subsection{Comparison to simulations}

\noindent
We compare the predictions of the ergodic model to simulations of the
late stages of terrestrial planet formation carried out by
\cite{hm13}\footnote{Many other authors have also simulated the
  giant-impact stage of planet formation
  \citep{cw98,acl99,cha01,kki06,rql06,mssm08,ray09,il10,kg10,mor10,fc14,pfy15}
  but the Hansen \& Murray simulation offers the most direct
  comparison to our model.}. They divide $20M_\oplus$ of material,
distributed between $0.05\au$ and $1\au$, into $\sim 30$--$40$ planets
and follow the evolution of these planets for 10 Myr. Collisions are
assumed to result in mergers. The calculation is repeated 100 times
with randomly varying initial conditions to build up the
statistics. Following Hansen \& Murray, we do not include planets with
$a>1.1\au$ in our analyses since these are mostly scattered objects
that are not expected to fit the ergodic model.

\begin{figure}[ht!]
\begin{center}
\includegraphics[width=0.8\textwidth]{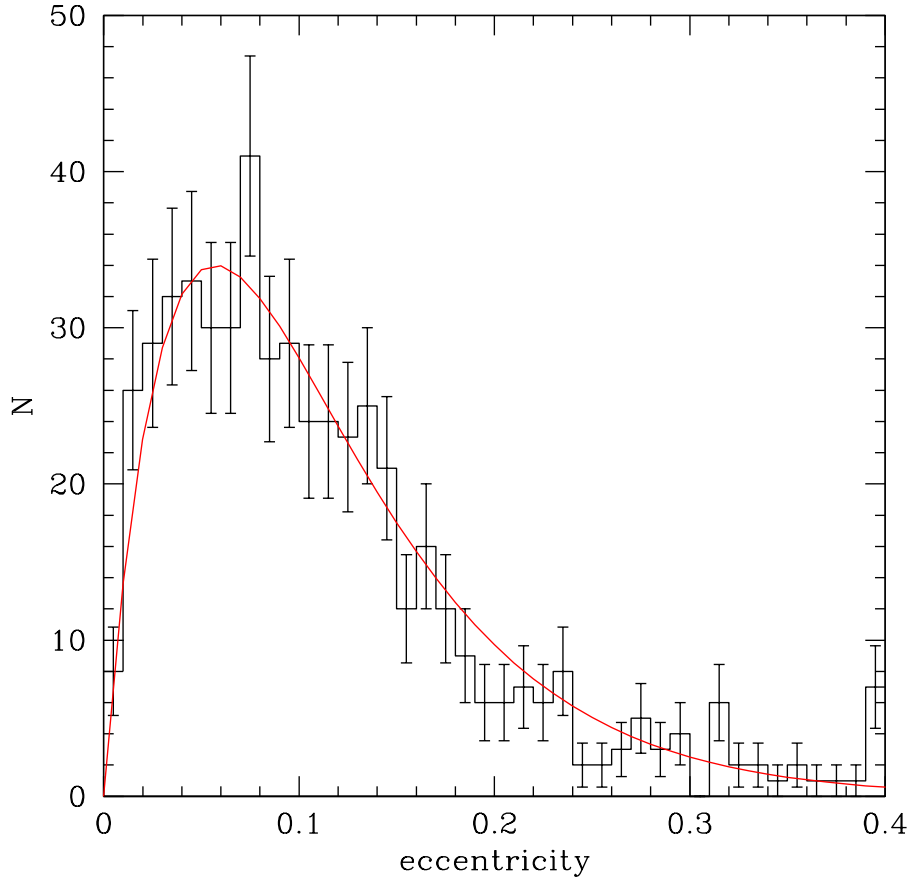}
\vspace{-1.0in}
\caption{\small The distribution of planetary eccentricities in the
  simulations of \cite{hm13}. The curve shows the prediction
  of equation (\ref{eq:beta}) with $\tau=0.057$.}
\label{fig:one}
\end{center}
\end{figure} 

Fitting equation (\ref{eq:beta}) to the eccentricity distribution of
527 surviving planets in the simulations we obtain $\tau=0.057\pm
0.002$ (1--$\sigma$ error, or reduction in log likelihood of
$\half$). With this value of $\tau$ the theoretical eccentricity
distribution (\ref{eq:beta}) is a very good match to the distribution
found in the simulations, as shown in Figure \ref{fig:one}. Indeed,
\cite{hm13} proposed the eccentricity distribution (\ref{eq:beta}) as
an {\em empirical} fitting formula for their results.

\begin{figure}[ht!]
\begin{center}
\includegraphics[width=0.8\textwidth]{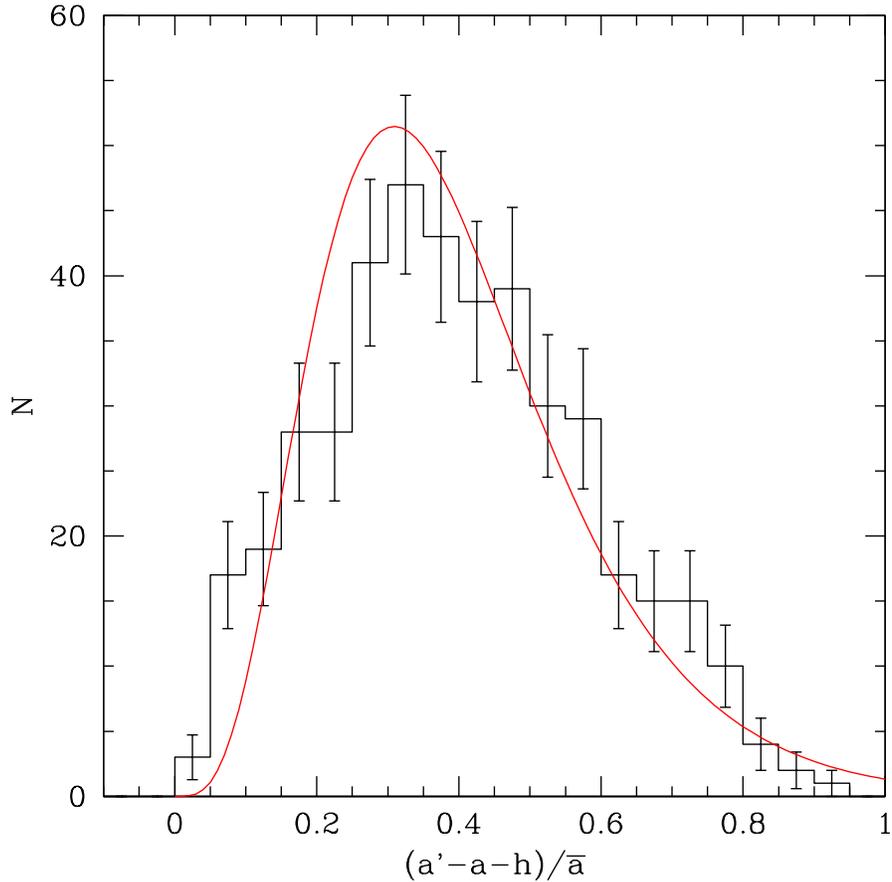}
\vspace{-1.0in}
\caption{\small The distribution of semimajor axis differences (minus
  the excluded length) in the simulations of \cite{hm13}. The curve
  shows the prediction of equation (\ref{eq:da}) with $\tau=0.067$.}
\label{fig:two}
\end{center}
\end{figure} 

Equation (\ref{eq:taufind}) relates the mean of the relative semimajor
axis differences of nearest-neighbor pairs to $\tau$. In this case the
relation depends on the stability criterion, i.e., on the excluded
length $h$. Using the stability criterion (\ref{eq:stab_ecc}) with
$\delcrit=9\pm 1$ we obtain $\tau=0.067\pm0.003$ for the 426 pairs in
the Hansen \& Murray simulation; the alternative stability criterion
(\ref{eq:cristobal}) yields $\tau=0.060$. These values for the
dynamical temperature are roughly consistent with the value determined from the
eccentricities, which is an encouraging test of the consistency of the
ergodic model. Moreover the distribution of semimajor axis differences
is fit well by the predicted distribution (\ref{eq:da}), as
shown in Figure \ref{fig:two}.  The filling factor, as determined by
equation (\ref{eq:ff}), is $F=0.30\pm0.03$ using the stability
criterion (\ref{eq:stab_ecc}) and 0.36 using (\ref{eq:cristobal}). 

\begin{figure}[ht!]
\begin{center}
\includegraphics[width=0.8\textwidth]{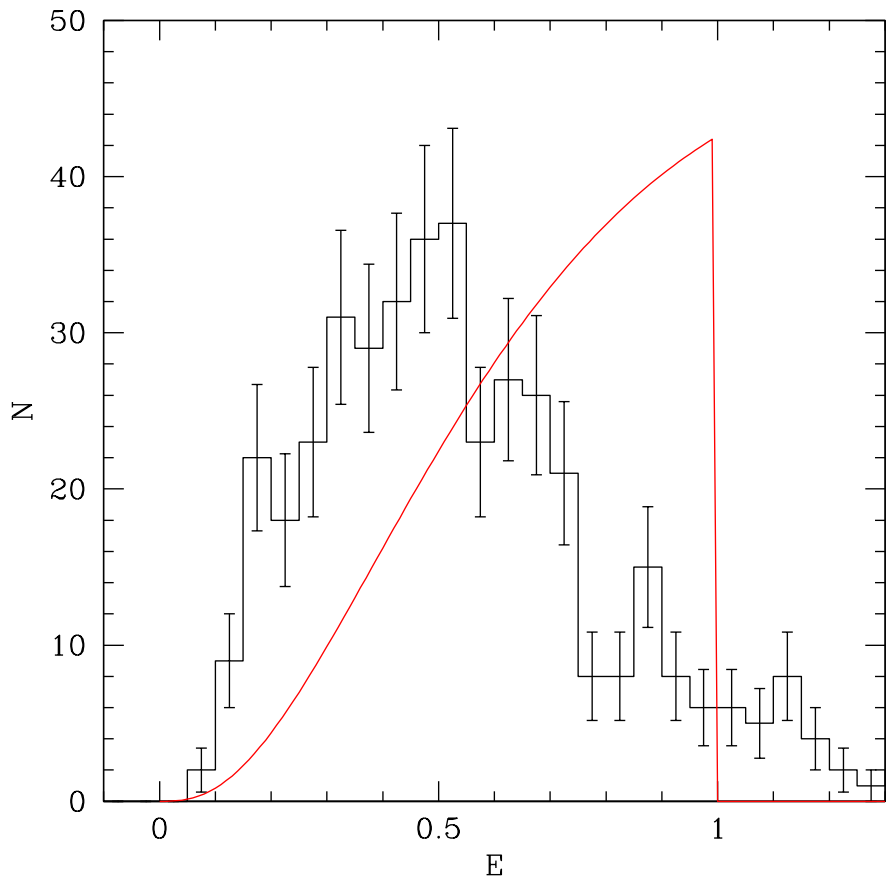}
\vspace{-1.0in}
\caption{\small The distribution of normalized eccentricity  (eq.\
  \ref{eq:enorm}) in the simulations of \cite{hm13}. The curve
  shows the prediction of equation (\ref{eq:enormpred}).}
\label{fig:two_a}
\end{center}
\end{figure} 

The distribution of normalized eccentricity $E$ (eq.\ \ref{eq:enorm})
in the simulations is shown in Figure \ref{fig:two_a}, along with the
prediction of the ergodic model. In this case the predicted
distribution does not match the simulations well. Part of the
discrepancy is that 11\% of the planet pairs have $E>1$, and these
would be unstable according to equation \ref{eq:hill}. The other major
discrepancy is that there are fewer planets in the simulation with
$E\gtrsim 0.6$--0.7 than the model predicts. These probably reflect two
oversimplifications of the ergodic model: (i) the stability criterion
is not a sharp boundary, as assumed in the derivation; (ii) planets
diffuse in eccentricity towards the stability boundary, so we expect
their phase-space density to be lower near the boundary that the
uniform density predicted by the ergodic model. 

\begin{figure}[ht!]
\vspace{1.0cm}
\begin{center}
\includegraphics[width=0.8\textwidth]{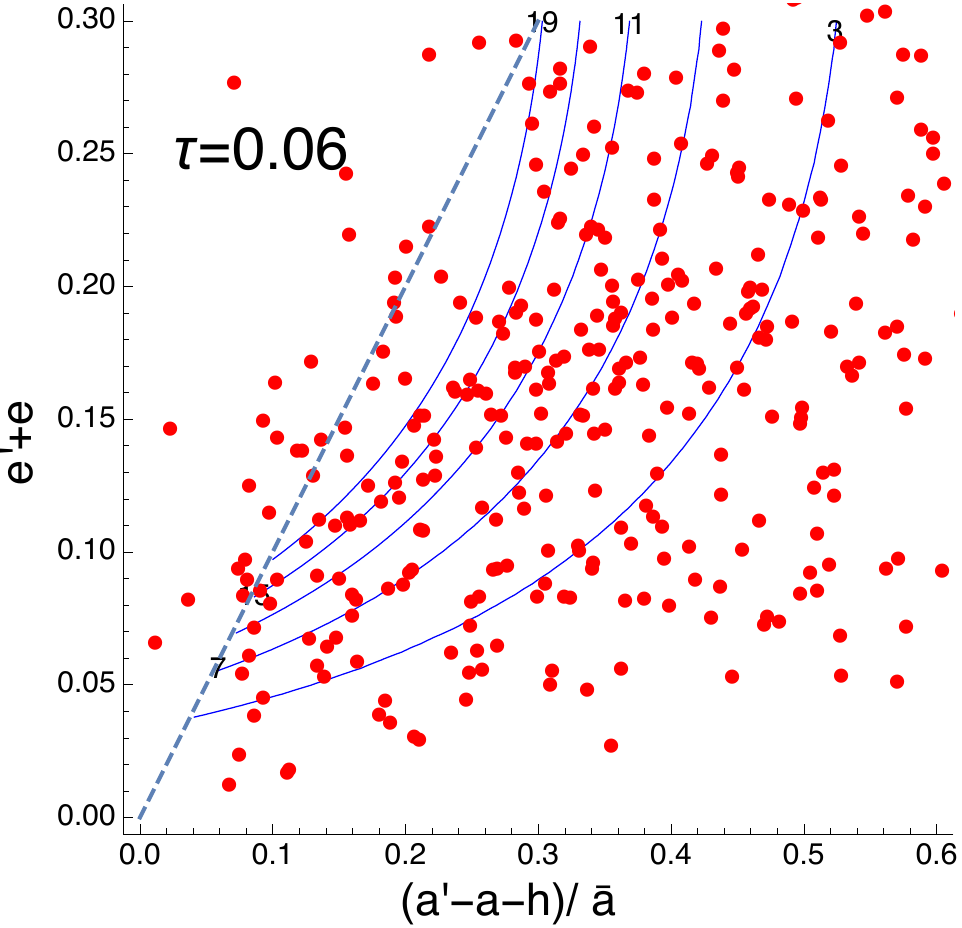}
\caption{\small Contours of the probability distribution for
  $\tau=0.06$ of the reduced semimajor axis difference distribution
  $a'-a-h$ and the total eccentricity (eq.\ \ref{eq:pdaet}), along
  with the values of these parameters for the nearest-neighbor pairs
  in the \cite{hm13} simulation. The labels on each contour give the
  relative probability density. The region above the dashed line is
  not allowed by the stability criterion (\ref{eq:hill} although a
  handful of simulated planet pairs are found there.}
\label{fig:cont}
\end{center}
\end{figure} 

Figure \ref{fig:cont} shows contour plots of the joint probability
density for the reduced semimajor axis difference $a'-a-h$ and total
eccentricity $e+e'$ of adjacent planets, derived from equation
(\ref{eq:pdaet}). The dynamical temperature is assumed to be
$\tau=0.06$. The red circles show the nearest-neighbor pairs in the
\cite{hm13} simulation. In this plot, contours of constant $E$ are
straight lines through the origin; the dashed line is $E=1$. The
deficit of planets at $E\gtrsim 0.6$--0.7 seen in Figure
\ref{fig:two_a} is evident, since the density of red circles does not
rise as fast as the probability density near the $E=1$ line. This
discrepancy is less visible in plots of the eccentricity or semimajor
axis distribution because these are projections onto the vertical or
horizontal axes. 

\subsection{Comparison to data}

\label{sec:data}

\noindent
The final and most important step is to compare our predictions to the
actual properties of exoplanets. There are several obstacles to
accomplishing this task: (i) Generally, eccentricities are only
available for planets whose orbits have been measured from radial
velocities. (ii) The masses of most planets discovered by Kepler can
only be estimated using an empirical mass-radius relation, and
individual planet masses exhibit large deviations from the mean
relation \citep{wm14}. (iii) There may be undiscovered planets in
between the known members of a multi-planet system and these would
affect the distribution of semimajor axis differences.

\paragraph{Kepler planets:} We have queried the NASA Exoplanet Archive
for all systems containing more than one confirmed planet discovered
by Kepler. This sample provides 932 planets in 362 distinct systems,
containing 556 nearest-neighbor pairs. We estimate the planetary
masses from the radii using the mass-radius relation from \cite{wm14}.
We then compute the relative semimajor axis difference for each pair,
$(a'-a)/\bar{a}$, using $\bar{a}=\half(a+a')$. Using the estimated
masses and the stability criterion (\ref{eq:hill}) with
$\delcrit=11\pm1$ we also compute the exclusion length $h$ and thus
determine the distribution of $(a-a'-h)/\bar{a}$, which ought to be
described by equation (\ref{eq:da}).

\begin{figure}[ht!]
\begin{center}
\includegraphics[width=0.8\textwidth]{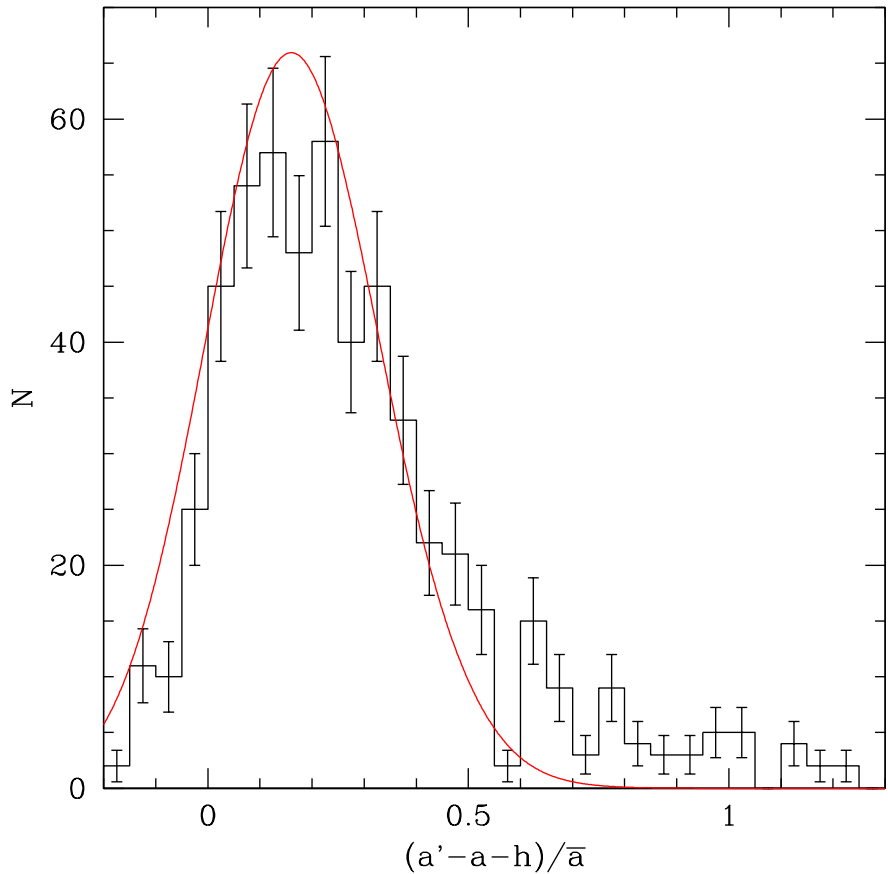}
\vspace{-1.0in}
\caption{\small The distribution of semimajor axis differences (minus
  the excluded length) for confirmed multiple-planet systems observed
  by Kepler. The curve shows the prediction of equation
  (\ref{eq:da}) with $\tau=0.028$, after convolving with a Gaussian
  having dispersion $\sigma_h=0.12$ in the relative semimajor axis difference to
  account for errors in the planetary masses.}
\label{fig:three}
\end{center}
\end{figure}

The result is shown as the histogram in Figure \ref{fig:three}. The
general shape of the histogram is similar to that of the histogram in
Figure \ref{fig:two} from the \cite{hm13} simulations. One obvious
difference, however, is that the histogram derived from the Kepler
data has a significant number of nearest neighbor pairs with negative
values of $a'-a-h$ (48 out of 556). These pairs should be unstable so
their presence must be explained. One possibility is that our
stability criterion is too conservative, but this is unlikely since
the analogous distribution from the \cite{hm13} simulations contains
no pairs with $a'-a-h<0$. Using the mass-radius relation from
\cite{fab14} or the alternative stability criterion
(\ref{eq:cristobal}) does not remove this difficulty: between 31 and
63 planet pairs in the sample are still apparently unstable. The most
plausible explanation is that the unstable pairs arise because of
scatter in the mass-radius relation.

To explore this possibility we shall assume that the distribution of
errors in the excluded length $h/\bar{a}$ (eq.\ \ref{eq:hill}) is
Gaussian, with standard deviation $\sigma_h$. Since $h/\bar{a}\propto
(m+m')^{1/3}$ for fixed stellar mass, $\sigma_h$ should be related to the error in $m^{1/3}$,
roughly as $\sigma_h/\langle h\rangle=\sigma_{m^{1/3}}/\langle
m^{1/3}\rangle$. To estimate the latter quantity, we use the sample of
65 exoplanets with measured masses and radii compiled by \cite{wm14},
and compare the observed values of $m^{1/3}$ with the predictions from
the mean mass-radius relation derived in that paper. (The mean mass of
the Weiss \& Marcy sample, $\langle(m/M_\oplus)^{1/3}\rangle=1.68$,
is close to the mean for the planets in our sample,
$\langle(m/M_\oplus)^{1/3}\rangle=1.71$, suggesting that the two
samples have similar properties.) We find $\sigma_{m^{1/3}}/\langle
m^{1/3}\rangle=0.41$. The mean excluded length in our sample is
$\langle h\rangle=0.25$ so we estimate $\sigma_h=0.10$ from these
arguments. 

We may now convolve our predicted distribution of semimajor axis
differences (\ref{eq:da}) with a Gaussian of width
$\sigma_h$ to find the distribution that would be obtained given
realistic errors in the planet masses. We fit the resulting
distribution to the data in Figure \ref{fig:three} using the
measurement error $\sigma_h$ and $\tau$ as fitting parameters, and
find a best fit with $\sigma_h=0.12\pm0.01$, $\tau=0.028\pm0.004$.
The best-fit value for $\sigma_h$ is remarkably close to the value
obtained from the analysis of the \cite{wm14} sample, $\sigma_h=0.10$;
thus the distribution of semimajor axis differences in the Kepler
sample appears to be consistent with the ergodic theory once the scatter
in the mass-radius relation is taken into account. 

The fit to the data has $\chi^2$ per degree of freedom
is 2.7, with most of the contribution to $\chi^2$ coming from a tail
of planets with $(a'-a-h)/\bar{a}\gtrsim 0.5$ that is not present in
the theoretical curve. The tail could arise from systems in which an
intermediate planet was below the detection limit of the Kepler survey.

The filling factor is $F=0.47\pm0.04$. Using the mass-radius relation
from \cite{fab14} instead of \cite{wm14} changes this by less than the
error bar, to $F=0.49$. Using the stability criterion
(\ref{eq:cristobal}) reduces the filling factor somewhat, to $F=0.42$.

Given the measured value $\tau=0.028\pm0.004$ for the Kepler semimajor
axis differences, the mean eccentricity should be $\langle
e\rangle=2\tau=0.05$--0.06 (eq.\ \ref{eq:beta}), unless the
eccentricities have been damped after the giant-impact phase is
complete \citep{hm15}. So far, we have only
limited information on the distribution of eccentricities of the
Kepler planets. (i) \cite{hl14} have estimated the eccentricities
using transit timing variations\footnote{There is a typographical
  error in the abstract of this paper. The quantity
  $0.018^{+0.005}_{-0.004}$ is not the rms eccentricity; it is
  $\sigma_e$ which equals the rms eccentricity divided by $\surd 2$.}
and obtain $\langle e\rangle=0.023\pm0.005$; for planets larger than
$2.5R_\oplus$ the mean eccentricity is a factor of two smaller, or
about $0.01$. (ii) \cite{mor11} have estimated eccentricities from the
distribution of transit durations and obtain $\langle
e\rangle=0.1$--0.25. The larger value relative to Hadden \& Lithwick
may arise in part because transit timing variations can only be
measured in multi-planet systems and such systems are expected to have
smaller eccentricities (see below); or because planets with large
transit timing variations are mostly near strong resonances, and such
planets could have a different eccentricity distribution. (iii)
Several authors have estimated the mean inclination of the Kepler
planets. \cite{dt11} find $\langle I\rangle <5^\circ$; in most
astrophysical disks, $\langle e\rangle=1$--2 times $\langle I\rangle$
so this result implies $\langle e\rangle < 0.17$. Similarly,
\cite{fm12} and \cite{fig12} find $\langle I\rangle <2^\circ$, so
$\langle e\rangle < 0.07$; \cite{joh12} find $\langle I\rangle
<2.5^\circ$, so $\langle e\rangle < 0.09$; and \cite{fab14} find
$\langle I\rangle \simeq 1.7^\circ$ and $\langle e\rangle\simeq\langle
I\rangle$ so $\langle e\rangle \simeq 0.03$. A concern with all of
these comparisons is that the multi-planet systems observed by Kepler
may be biased towards low eccentricities because eccentricity and
inclination are correlated and low-inclination systems are more likely
to have multiple transits. We conclude that the observations so far
are roughly consistent with the estimate of the mean eccentricity from
the ergodic model, $\langle e\rangle=2\tau=0.05$--0.06, but do not
provide strong support for it.

The filling factor estimated from the \cite{hm13} simulations is
$F=0.36\pm0.03$, while the filling factor estimated from the semimajor
axis differences of Kepler planets is 0.4--0.5 depending on the
mass-radius relation and stability criterion. If the dynamics of the
late stages of formation of the Kepler planets are faithfully modeled
by the simulations, we might expect the two filling factors to be the
same. The similarity of the two numbers is impressive but it is
worthwhile to ask why they might differ. One intriguing possibility is that the
Kepler planet masses have been overestimated; decreasing the masses by
a factor of two would decrease the Kepler filling factor by about 0.1
and bring it to agreement with the filling factor in the simulations. 

The values of the dynamical temperature $\tau$ are also similar: 0.06
in the \cite{hm13} simulations and 0.03 in the Kepler data. This
difference cannot be accounted for by differences in the surface
density or mass: the scalings described after equation
(\ref{eq:enormpred}) imply $\tau\sim \langle m^{1/3}\rangle$ and
$\langle (m/M_\oplus)^{1/3}\rangle_{\rm HM}/\langle (m/M_\oplus)^{1/3}\rangle_{\rm Kepler}=1.45/1.71=0.85$, 
which would predict that $\tau$ should be 15\% smaller in the
simulations than in the data. A more likely
cause of the difference is that the simulations did not allow for
fragmentation: the arguments following equation (\ref{eq:tauup}) imply
that if fragmentation is present the dynamical temperature $\tau$
cannot exceed 0.02--0.03, consistent with the Kepler data.

\paragraph{Radial-velocity planets:} We have queried the Exoplanets
Data Explorer \citep{han14} for multiple-planet systems discovered by
radial-velocity variations in their host stars. This sample provides 135
planets in 55 systems or 77 nearest neighbor pairs. The distribution
of eccentricities is fit well by equation (\ref{eq:beta}) with
$\tau=0.10$; this value exceeds the upper limit derived after equation
(\ref{eq:tauup}) by a factor of three or more but this is not a
contradiction because the giant planets probably did not grow to their
present masses by giant impacts. The distribution of semimajor axis
differences, however, is not well-fit by the ergodic model. There are
several likely reasons for this. (i) Almost one-third of the planet
pairs violate the stability criterion (\ref{eq:stab_ecc}) (we
estimated the masses assuming the systems are edge-on, which gives a
lower limit); this is partly because the criterion is not valid for
planets in mean-motion resonances, which are common among giant
planets. (ii) The ergodic model is based on the sheared-sheet
approximation, which requires that the semimajor axis difference
$a'-a\ll \bar{a}$; this is generally not true for planets larger than
a Jupiter mass since the excluded length $h$ (eq.\ \ref{eq:hill})
satisfies $h/\bar{a}=0.96(\delcrit/11)[(m+m')/2M_{\rm
  Jupiter}]^{1/3}$. (iii) Giant planets formed by processes that
differ from those of terrestrial planets, for example migration and
the accretion of massive gas envelopes. As described at the end of
\S\ref{sec:introd}, the ergodic model is only expected to apply to
giant planets if their orbits are mostly determined by late-stage
dynamical evolution.

Equation (\ref{eq:beta})  suggests that in an ensemble of systems
with the same filling factor the mean eccentricity should vary as
$1/N$ (more precisely, inversely with the number of planets per unit
radius). \cite{lim14} find that for known exoplanet systems the
eccentricity decreases with multiplicity roughly as $e\sim N^{-1.2}$,
in reasonable agreement with this prediction.

\begin{figure}[ht!]
\begin{center}
\includegraphics[width=0.48\textwidth]{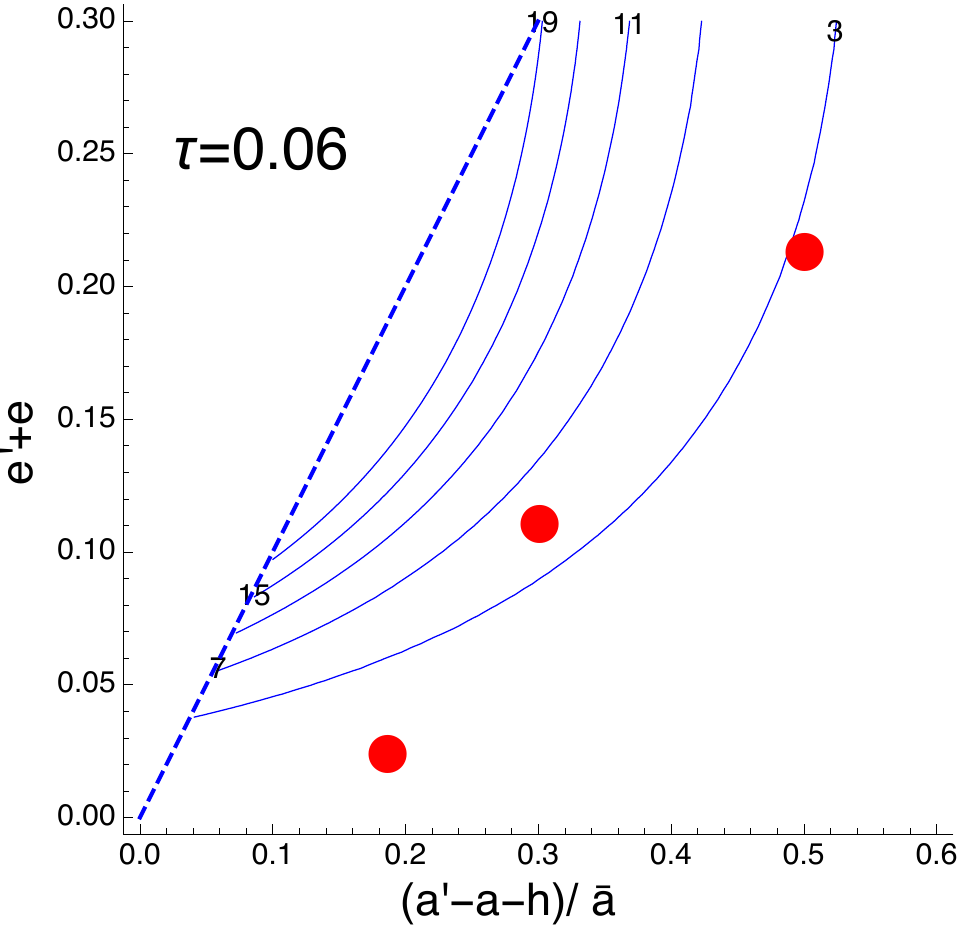}
\includegraphics[width=0.48\textwidth]{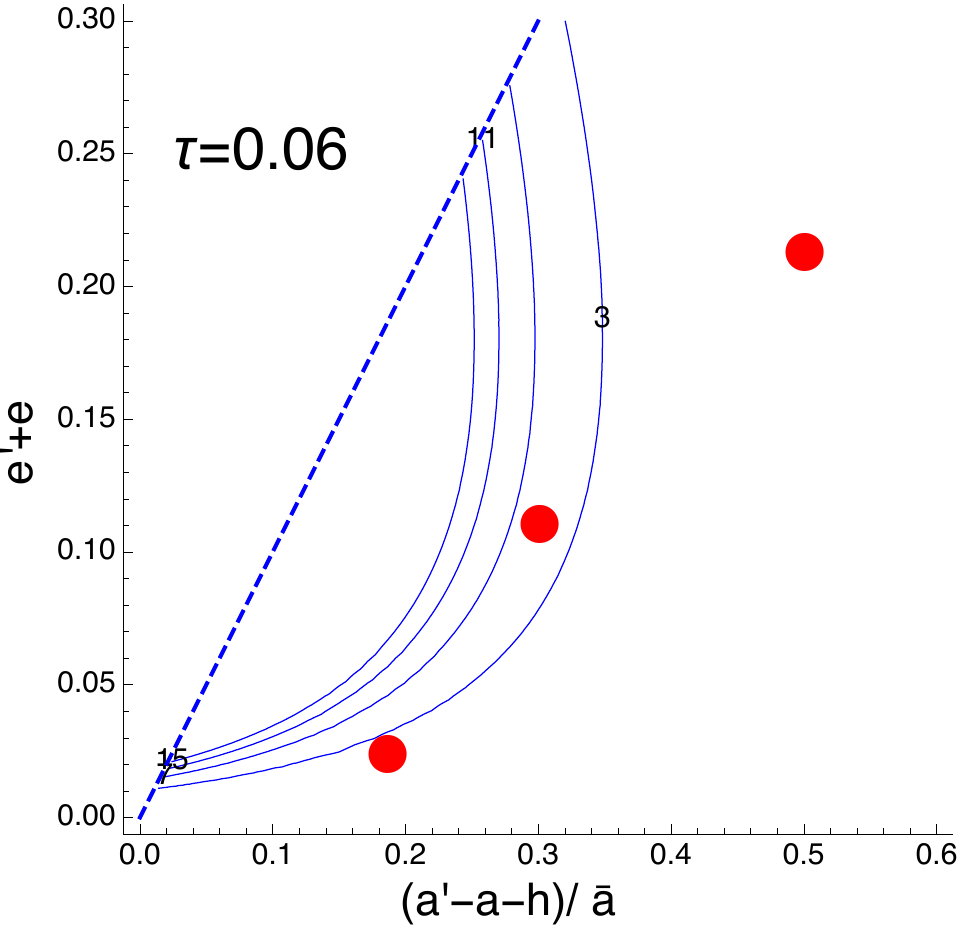}
\caption{\small As in Figure \ref{fig:cont}, except the planet pairs
  plotted are Mercury+Venus, Venus+Earth, Earth+Mars. The left plot is
  for dynamical temperature $\tau=0.06$ and the right for
  $\tau=0.03$. }
\label{fig:four}
\end{center}
\end{figure} 

\paragraph{The solar system:} Figure \ref{fig:four} shows contour
plots of the joint probability density for the reduced semimajor axis
difference $a'-a-h$ and total eccentricity $e+e'$ of adjacent planets,
derived from equation (\ref{eq:pdaet}). The dynamical temperature
is assumed to be $\tau=0.06$ in the left panel, the same as we found for the Kepler
planets, and 0.03 on the right. The red circles show the actual values
for the three nearest-neighbor pairs among the four terrestrial
planets. For $\tau=0.06$ most of the weight of the probability density
distribution lies at total eccentricities larger than those of the
solar-system planet pairs. For $\tau=0.03$ the Mercury-Venus pair
has too large a semimajor axis difference relative to the probability
density distribution. These mismatches presumably
reflect the well-known difficulty that simulations of the giant-impact
phase produce eccentricities for the terrestrial planets of the solar
system that are too large, unless there is a residual population of
small planetesimals to damp the eccentricities \citep[e.g.,][]{mor12}.

The outer planets of the solar system do not fit the ergodic model
well, in part because Jupiter and Saturn are separated by only eight
mutual Hill radii and so would nominally be unstable according to our
crude stability criterion.

\section{Discussion}

\label{sec:disc}

\noindent
We have described a simple model for the distribution of semimajor
axes and eccentricities of planets. The model assumes that the last
phase of terrestrial planet formation was the giant-impact phase, and
is based on the simple ansatz that planets are uniformly distributed
over the volume of phase space in which their orbits are stable for
the lifetime of the planetary system (the ``ergodic model''). Our
model yields predictions, in terms of a single free parameter $\tau$
that we call the dynamical temperature, for the distribution of
eccentricities (eq.\ \ref{eq:beta}), the distribution of separations
of nearest neighbors (eq.\ \ref{eq:da}), and more generally for any
property derivable from the complete $N$-planet distribution
function. For example, the ergodic model predicts that in a given
system the eccentricities should be independent of planetary
mass. N-body simulations generally report only a weak negative
correlation between eccentricity and mass \citep[e.g.,][]{cha13}.

The ergodic model has many limitations.  (i) It does not account
for the likely influence of giant planets at larger radii on the
formation of terrestrial planets. (ii) Different planetary systems may
have different dynamical temperatures $\tau$ and fitting the data from
a large ensemble of systems to the ergodic model will only work well if most
systems have similar temperatures. (iii) Our analysis is based on the
sheared-sheet or Hill's approximation and hence does not work well
when the planetary masses are large enough that the relative
separations $(a'-a)/\bar{a}$ or the fractional excluded lengths $h$
(eq.\ \ref{eq:hill}) are of order unity or larger; typically this
occurs for planets more massive than Jupiter. (iv) The stability
criteria (\ref{eq:hill}) and (\ref{eq:cristobal}) are only approximate
and probably no simple criterion perfectly separates stable from
unstable orbits in a multi-planet system \citep{pet15}. (v) Our
assumption that the orbits are distributed uniformly over the stable
part of phase space is shaky, essentially because the neighboring
unstable regions represent an absorbing barrier rather than a
reflecting one. A better approximation would require following the diffusion
of planet orbits through phase space, but implementing this would
require a reliable model for the rate of this diffusion and how it
depends on the orbits of the planet and its neighbors. 

In this paper we examine only the distribution of orbital elements and not the
distribution of planetary masses that emerges in the giant-impact
phase, and any complete statistical model of this phase should predict
both. An  interesting question is whether the system
described in \S\ref{sec:system}, based on the sheared-sheet or Hill's
approximation, exhibits long-range order in the masses as
$N\to\infty$; in other words if the radial width $\Delta a\sim N$ but
the average surface density is fixed, does the mass of the most
massive planet grow as $m_{\rm max}\sim N^k$ and if so what is the
critical exponent $k$?

Although the ergodic model is simple and physically plausible, there
are other approaches to the statistical mechanics of planet
formation. An interesting alternative is due to
\cite{las00}, who pointed out that in the secular approximation there
is an important conserved quantity: the  angular-momentum
deficit, 
\begin{equation}
C=\sum_{i=1}^N (GM_\star a_i)^{1/2}[1-(1-e_i^2)^{1/2}\cos I_i],
\label{eq:amd}
\end{equation}
that is, the difference between the total angular momentum of the
planets and the angular momentum that they would have on circular,
coplanar orbits with the same semimajor axes. He hypothesized that the
eccentricities and inclinations of planets evolve randomly, subject to
conservation of the angular-momentum deficit, until there is a binary
collision; in each collision the angular-momentum deficit is reduced;
and collisions cease and the system becomes permanently stable once
the angular-momentum deficit becomes too small to allow any more close
encounters. Laskar's model is not unique, since it requires an {\em ad
  hoc} prescription for the ``random'' evolution of the orbits;
however, once this prescription is implemented it is straightforward
to predict both the masses and the orbital elements of the planets in
an ensemble of systems. Laskar's model differs from ours in that
it generally predicts a strong anti-correlation between eccentricity
and mass. 

One-dimensional models are powerful tools in statistical mechanics
because they are often much simpler to solve than their
three-dimensional analogs \citep{lieb66}. It is remarkable that the
giant-impact phase of planetary formation is most naturally modeled in
one physical dimension, radius, basically because (i) the systems are
nearly flat; (ii) orbital and apsidal motion effectively averages the
orbital and collisional dynamics over the azimuthal angle. Although
the ergodic model contains many simplifications, more general and
accurate analyses of the statistical mechanics of the giant-impact phase
could still be one-dimensional in this sense and therefore amenable to
analytic treatments. 

Exact models in statistical mechanics are mainly useful because they
provide insight into the behavior of real systems, rather than because
they are accurate representations of them. Similarly, the ergodic
model presented here is mainly useful because it provides simple
predictions for many properties of the 1--, 2--, and even $N$--planet
joint distribution of orbital elements. These predictions
encapsulate some, though certainly not all, of the physics of the late
stages of planet formation, and therefore should help to organize and
interpret both observations of planet orbits and numerical simulations
of planet formation.

This research was initially stimulated by conversations with Renu
Malhotra. I thank Brad Hansen for comments on the manuscript, and for
providing the results from his simulations. I am particularly grateful
to Cristobal Petrovich for discussions, insight, and pointers to the
literature, and for carrying out exploratory N-body integrations on my
behalf.

\end{document}